\documentclass[letterpaper, 10 pt, conference]{ieeeconf}  % Comment this line out if you need a4paper
\usepackage{graphicx}
\usepackage{amsmath,amsfonts,mathrsfs,amssymb,dsfont}
\usepackage{booktabs}
\usepackage{cite}

\IEEEoverridecommandlockouts                              % This command is only needed if 
\title{\LARGE \bf
Moment Closure Approximations in a Genetic Negative Feedback Circuit
}

\author{Mohammad Soltani$^{1}$, Cesar Vargas$^{1}$, Niraj Kumar$^{2}$, Rahul Kulkarni$^{2}$, Abhyudai Singh$^{1}$% <-this % stops a space
\thanks{$^{1}$M. Soltani, C. Vargas and A. Singh are with Department of Electrical and Computer Engineering, Newark, DE USA 19716.
{\tt\small msoltani@udel.edu, cavargar@udel.edu, absingh@udel.edu}}%
\thanks{$^{2}$N. Kumar and R. Kulkarni are with Department of Physics, University of Massachusetts Boston, Boston, MA USA 02125.
        {\tt\small Niraj.Kumar@umb.edu, rahul.kulkarni@umb.edu}}%
}

\begin{document}

\maketitle
\thispagestyle{empty}
\pagestyle{empty}

%%%%%%%%%%%%%%%%%%%%%%%%%%%%%%%%%%%%%%%%%%%%%%%%%%%%%%%%%%%%%%%%%%%%%%%%%%%%%%%%
\begin{abstract}
Auto-regulation, a process wherein a protein negatively regulates its own production, is a common motif in gene expression networks. Negative feedback in gene expression plays a critical role in buffering intracellular fluctuations in protein concentrations around optimal value. Due to the nonlinearities present in these feedbacks, moment dynamics are typically not closed, in the sense that the time derivative of the lower-order statistical moments of the protein copy number depends on high-order moments. Moment equations are closed by expressing higher-order moments as nonlinear functions of lower-order moments, a technique commonly referred to as moment closure. Here, we compare the performance of different moment closure techniques. Our results show that the commonly used closure method, which assumes a priori that the protein population counts are normally distributed, performs poorly. In contrast, conditional derivative matching, a novel closure scheme proposed here provides a good approximation to the exact moments across different parameter regimes.  In summary our study provides a new moment closure method for studying stochastic dynamics of genetic negative feedback circuits, and can be extended to probe noise in more complex gene networks.
\end{abstract}
%%%%%%%%%%%%%%%%%%%%%%%%%%%%%%%%%%%%%%%%%%%%%%%%%%%%%%%%%%%%%%%%%%%%%%%%%%%%%%%%
\section{INTRODUCTION}

The stochastic nature of the gene expression process creates considerable random fluctuations in protein levels over time inside individual living cells \cite{bpm06,rao05,Li:2010uv,arm98,paun04,pau05,els02,bkc03}. Noise in protein levels corrupt information processing in gene networks \cite{lps07}, and is detrimental for the functioning of essential proteins whose levels have to be maintained within certain bounds for optimal performance \cite{lps07,fhg04,leh08}. Not surprisingly, cells use a variety of regulatory mechanisms to buffer stochasticity in protein levels \cite{sak06,sih09c,lvp10,bhj03,pep08,moa04,swa04}. The most common and simplest example of such a mechanism is auto-regulation, wherein proteins expressed from a gene inhibit their own synthesis \cite{alo07,whs03,bhs08,thp98,sih09b}. Here we develop approximate methods to study stochastic dynamics of auto-regulatory genetic circuits.

Nonlinear propensity functions in these negative feedback systems lead to the well-known problem of moment closure: time derivative of the lower-order statistical moments of the protein copy number depends on high-order moments. Moments are typically solved by performing moment closure, which closes the differential equations by expressing higher order moments as functions of lower order moments. Various closure techniques have recently been proposed to study noise in the biochemical systems \cite{gov06,lkk09,gou05,sih10,gillespie2009}. The goal of this study is to test existing and new moment closure methods in the their ability to capture 
stochasticity in auto-regulatory gene networks.  

Exact moment dynamics are generally computed by running a large number of Monte Carlo simulations of the gene network of interest \cite{GillespiePetzoldOct03,gil01}. However, it turns out that for 
auto-regulatory gene networks, exact closed-form solutions for the protein moments can be obtained under certain assumptions of short mRNA half-life and non-cooperative feedback. These exact formulas are used to benchmark the performance of different moment closure techniques.  Our analysis reveals poor performance of existing closure methods. In contrast, conditional derivative matching, a new closure technique proposed in this study provides moment that are remarkably close to the exact solution for a wide range of parameter values.

The paper is organized as follows: stochastic model of an auto-regulatory gene is introduced in Section II. Exact solution of the model is provided in Section III. Moment dynamics of the negative feedback system are obtained in Section IV, and closed using different moment closure techniques in Section V. Performance of closure methods are compared in Section VI. Finally, conclusions and direction of future work are discussed in Section VII.

\section{Model Description}
Model schematic of a self-regulating gene is illustrated in Fig. 1. The gene can reside in two possible states: a transcriptionally active (ON) and inactive (OFF) state, with mRNA production only occurring from the ON state.
Let $g(t)$ be a Bernoulli
random variable with $g(t)=0$ ($g(t)=1$) denoting that the gene is active (inactive) at time $t$. An approximation often used to simplify gene expression models is that the mRNA half-life is considerably shorter than the protein half-life. In this physiologically relevant parameter regime, 
one can ignore mRNA dynamics and model protein production as bursty birth-death process \cite{shs08,sih09b}. Towards that end, we assume that 
protein bursts occur at a rate $k_p$ when $g(t)=1$. Consistent with data \cite{gpz05}, each burst generates
$B$ protein molecules, where $B$ is a geometrically distributed random variable with distribution
\begin{equation}
        \begin{aligned}
&{\rm Probability}\{B=i\}=\alpha(i)=(1-s)^is, \\
& \hspace{35mm} 0<s\leq1, \ \   i=\{0,1,2,\ldots\}.
        \end{aligned}
\label{1111}
\end{equation}
The mean burst size is given by $\langle B \rangle:=(1-s)/s$, where $\langle . \rangle$ represents the expected value. Finally, each protein molecule decays with at constant rate $\gamma_p$.
  \begin{figure}[thpb]
      \centering
      \framebox{{\includegraphics[scale=1.9]{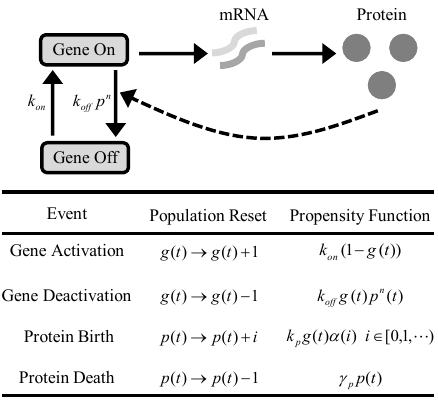}}}
      \caption{{\bf Schematic of a negative feedback loop in gene expression}. Proteins expressed form the active (ON) state of the gene turn the gene OFF. Stochastic model consists of four events
      that ``fire" probabilistically. Whenever an events occurs the state of the system resets based on the second column of the table. Third column lists the event propensity
      function, which determines how often the events occur. The state of the gene is denoted by $g(t)$, with $g(t)=1$ ($g(t)=0$) denoting that the gene is transcriptionally active (inactive) at time $t$.
      Protein copy number is represented by $p(t)$.}
      \label{figmodel}
   \end{figure}

To control expression levels many genes employ negative feedback loops, where protein molecules bind to their own gene promoter and block overproduction  \cite{alo06}.
This feedback is incorporated in the model by assuming that 
the gene transitions from the ON to the OFF
state with rate $k_{off}p^n(t)$, where $p(t)$ is the protein level in the cell at time $t$ and $n$ denotes the extent of cooperativity in the feedback system \cite{alo06}. 
After a gene becomes transcriptionally inactive, it turns ON again with rate $k_{on}$. Note that $k_{on}$ is inversely related to the protein binding affinity, with stronger binding resulting in more repression and lower values of $k_{on}$. In the limit $k_{on}\to\infty$, gene expression is constitutive (i.e., gene is always transcriptionally active) with no feedback regulation. 

Based on the standard stochastic formulation of chemical kinetics \cite{mcq67,gil01}, the model comprises of four events that occur probabilistically at exponentially-distributed time intervals (Fig. 1). The first two events correspond to gene activation/deactivation. We assume that protein levels are sufficiently large such that gene deactivation/activation (which occurs due to protein binding/unbinding to the promoter) does not significantly change $p(t)$. The last two events represent protein production in geometric bursts, and protein degradation. Whenever an events occurs, $g(t)$ and $p(t)$ are resets based on the second column of the table. Third column lists the event propensity function $f(g,p)$, which determines how often the reactions occur. In particular, the probability that an event occurs in the next
infinitesimal time interval $(t,t+dt]$ is $f(g,p)dt$.

An exact analytical solution of this model is generally not possible for any arbitrary $n$. However, as shown below, closed-form solutions of the statistical moments can be obtained
for $n=1$ (non-cooperative feedback) \cite{kpk14}. These solutions are later used to benchmark different moment closure methods. 
Note that even for $n=1$, the gene deactivation propensity function is nonlinear. 

\section{Exact Solution of Moments}
Let $P_
{1}(p,t)$ ($P_
{0}(p,t)$) denote the
probability that at time $t,$ the gene is in the active (inactive)
state with $p$ number of protein molecules inside the cell. Then, the probability of observing $p$ protein molecules at time $t$ is given by
\begin{equation}
P(p,t)=P_
{1}(p,t)+P_
{0}(p,t).
\end{equation}
For the stochastic model described in Fig. 1, these probabilities evolve according to 
the following Chemical Master Equation (CME):
\begin{figure*}[!t]
\setcounter{equation}{10}
\begin{small}
\begin{align}\tag{11}\label{moment_dynamics}
\frac{d}{dt}\left[\begin{array}{c}\langle g \rangle \\ \langle p \rangle \\ \langle gp \rangle \\  \langle p^2 \rangle \\ \langle gp^2 \rangle  \\ \langle p^3 \rangle  \end{array} \right] = \left[\begin{array}{c} k_{on} \\ 0 \\ 0 \\0 \\ 0 \\ 0  \end{array} \right]+\left[\begin{array}{cccccc} -k_{on} & 0 & -k_{off} &0 & 0& 0 \\
k_p \langle B \rangle & -\gamma_p & 0 & 0 & 0 & 0\\
k_p \langle B \rangle & k_{on} & -\gamma_p-k_{on} &0 & -k_{off} & 0\\
k_p \langle B^2 \rangle & \gamma_p & 2 k_p \langle B \rangle & -2\gamma_p & 0 & 0\\
k_p \langle B^2 \rangle & 0 & 2k_p \langle B \rangle +\gamma_p & k_{on} & -2\gamma_p-k_{on} & 0 \\
k_p \langle B^3 \rangle & -\gamma_p & 3 k_p \langle B^2 \rangle  & 3\gamma_p & 3 k_p \langle B \rangle & -3\gamma_p \\
\end{array} \right] \left[\begin{array}{c}\langle g \rangle \\ \langle p \rangle \\ \langle gp \rangle \\  \langle p^2 \rangle \\ \langle gp^2 \rangle  \\ \langle p^3 \rangle  \end{array} \right]+ \left[\begin{array}{c} 0 \\ 0 \\ 0 \\ 0 \\ -k_{off}  \\ 0 \end{array} \right]\langle gp^3 \rangle
\end{align}
\end{small}
\setcounter{equation}{2}
\end{figure*}
%
%Backward Kolmogorov equation shows characteristics of dynamic random systems. In engineering, backward Kolmogorov equation is known as chemical master equation. CME shows time evolution of the system between its states, and gives the probability of being in each state. Solution of CME can gives any statistical characteristics of the system.\\
%Negative feedback gene regulatory network is a dynamic random system where different states of the system are protein population and state of gene. Therefore CME can be solved for this system based on propensity functions and population resets in Figure ~\ref{figmodel}.\\
%Let us denote by $P_
{}
%g,p)$ as the probability that there are $p$ proteins at a time $t$ given that the gene is in the state $g$ with $g$ = 0 or 1 representing its inactive or active states, respectively. As we mentioned before, by denoting $\alpha(i)$ as the probability of $B=i$ proteins produced in a burst, the temporal evolution of the system is then given as \cite{kpk}
\begin{equation}
        \begin{aligned}
& \frac{\partial P_
{1}(
p,t)}{\partial t}=k_p\sum_{i=0}^p \alpha(i)P_
{1}(
p-i,t)\\
&\hspace{17mm} +\gamma_p (p+1)P_
{1}(
p+1,t) + k_{on} P_
{0}(
p,t)\\
&\hspace{17mm} -\left( k_p + k_{off}p + \gamma_p p\right) P_
{1}(
p,t),\\
& \frac{\partial P_
{0}(p,t)}{\partial t}= \gamma_p (p+1)P_
{0}(p+1,t) + k_{off} p P_
{1}(p,t)\\
&\hspace{17mm}-\left(k_{on}+\gamma_p p \right) P_
{0}(p,t).
        \end{aligned}
        \label{MasterEq1}
\end{equation}
Generating functions corresponding to $P_
{0}(p,t)$, $P_
{1}(p,t)$ and $P(p,t)$
are defined as
\begin{subequations}\label{01}
\begin{align}
&G_{0}(z,\, t):=\sum_{p=0}^\infty z^{p}
\, P_
{0}(p,t), \ G_{1}(z,\, t):=\sum_{p=0}^\infty z^{p}
\, P_
{1}(p,t) \\ 
& G(z,t):=G_{0}(z,\, t)+G_{1}(z,\, t)=\sum_{p=0}^\infty z^{p}
\, P(p,t). \label{eq:3}
\end{align}
\end{subequations}
Using \eqref{01}, the CME can be transformed into coupled PDEs for the corresponding generating functions $G_{0}(z,t)$ and $G_{1}(z,t)$. By carrying out a series of transformations, the generating function for the steady-state protein distribution $G(z)$ can be obtained as
\begin{equation}
G(z)=\frac{_2F_1[u,v;u+v+1-w;1-\phi\{1+\langle B \rangle(1-z)\}]}{_2F_1[u,v;u+v+1-w;1-\phi]}, \label{exact}
\end{equation}
where $_2F_1$ is the hypergeometric function and $u$, $v$, $w$ and $\phi$, are related to model parameters by
\begin{equation}
        \begin{aligned}
&u+v=\frac{k_p + k_{on}}{\gamma_p+k_{off}}\\
&uv=\frac{k_{on}k_p}{\gamma_p(\gamma_p+k_{off})},\\
& \phi = \frac{\gamma_p + k_{off}}{\gamma_p+k_{off} +\langle B \rangle k_{off}},\\
& w=\frac{k_p + \gamma_p + k_{off}(1+\langle B \rangle)}{\gamma_p + k_{off}(1+\langle B \rangle)}
        \end{aligned}
\end{equation}
 \cite{kpk14}. Once we have an explicit expression for $G(z)$, steady-state first and second-order statistical moments of the protein copy number are obtained as
\begin{align}\label{act}
\langle p \rangle=\frac{dG(z)}{dz}|_{z=1}, \ \
\langle p^2 \rangle =\frac{d^2G(z)}{dz^2}|_{z=1}+  \langle p \rangle.
\end{align}
In the following sections, statistical moments are computed using various moment closure techniques and compared to \eqref{act} to test their accuracy. We begin by deriving 
differential equations describing the time evolution of the \emph{uncentered} moments of $p(t)$ and $g(t)$.

\section{Computing Moment Dynamics}
Using the above CME it can be shown that for any function $\varphi(g,p)$,
\begin{equation}
        \begin{aligned}
& \dfrac{d \langle \varphi(g,p) \rangle}{dt}=
& \left \langle  \sum_{Events}  \Delta \varphi(g,p) \times f(g,p) \right \rangle,
        \end{aligned}
\label{eqn2_moments}
\end{equation}
where $\Delta \varphi(g,p)$ is the change in $\varphi$ when an event occurs, and $f(g,p)$ is the event propensity function. Moment dynamics are obtained by choosing $\varphi(g,p)$
to be an appropriate monomial of the form $g$, $p$, $gp$, $p^2$, and using resets/propensity functions in Fig. 1. \\

Let $\mu=[\langle g \rangle, \langle  p \rangle,\langle  gp \rangle, \langle p^2 \rangle, \langle gp^2 \rangle, \langle p^3 \rangle ]^T$ be a vector of all moment of $p(t)$ and $g(t)$ up to order three. Moments $\langle g^2 \rangle$, $\langle g^3 \rangle$ and $\langle gp^2 \rangle$ were not included in $\mu$ since for a Bernoulli random variable $g(t)$ 
\begin{align}
\langle g^{j}\rangle&=\langle g \rangle, \ j\in\{2,3,\dots\}.\\
\langle g^{j}p^k \rangle&=\langle gp^k \rangle, \ j,k \in\{1,2,3,\dots\}.
\end{align}
Using \eqref{eqn2_moments},
the time evolution of $\mu$ is given by \eqref{moment_dynamics} which can be compactly represented as a linear system
\setcounter{equation}{11}
\begin{align}
\dot{\mu}= \hat{a}+ A \mu + B \bar{\mu}\label{comp},
\end{align}
where $\bar{\mu}=\langle gp^3 \rangle$ is a fourth-order moment and vector $\hat{a}$, matrices $A$, $B$ depend on model parameters. The nonlinear propensity function leads to unclosed moment dynamics, where the time evolution of third-order moments depends on fourth-order moments. To solve \eqref{comp}, closed system of nonlinear differential equations are obtained 
by approximating $\bar{\mu}\approx \theta(\mu)$ as 
a nonlinear function of moments up to order three, in which case
\begin{align}
\dot{\mu}\approx \hat{a}+ A \mu + B \theta(\mu)\label{compa}.
\end{align}
We refer to $ \theta(\mu)$ as the \emph{moment closure function}. 

%

%are constant matrices. $\bar{\mu}$ is a vector of time varying higher order moments. Moment closure methods approximate $\bar{\mu}$ as a nonlinear function of lower order moments ($\bar{\mu}=\varphi(\mu)$).\\
%First, $\varphi$ was selected as $g,\hspace{1mm} p,\hspace{1mm} p^2,\hspace{1mm} gp$ and moment closure was used to close set of second order moments (Higher order term in this case was $gp^2$.). But for second order moments the error between approximated moments and exact solution was high. In order to decrease the error, third order moments are considered and approximation is done for them.\\
%One does not need to consider all the third order moments. $g$ just can be zero or one, therefore gene is a binary random variable. For a binary random variable higher order moments are equal to the first order moment ($\langle g \rangle = \langle g^{l_1} \rangle \hspace{10 pt} l_1=2,3,\ldots$). Thus dynamics of $\langle g^2 \rangle$ and $\langle g^3 \rangle$ are not needed. In addition, for binary random variable any joint moment will be simplified as $\langle gp^{l_2} \rangle = \langle g^{l_1} p^{l_2} \rangle  \hspace{10 pt}  \forall{l_1}>0$. Therefore $g^2p$ is not needed. Based on these facts, set of third order moments is complete by selection of $\varphi$ as $g,\hspace{1mm} p,\hspace{1mm} p^2,\hspace{1mm} gp,\hspace{1mm} gp^2,\hspace{1mm} p^3$. Moment dynamics are shown in equation \eqref{moment_dynamics}.
%\newpage

\section{Moment Closure}
Next, moment closure functions are derived based on four different closure methods: Gaussian approximation, Conditional Gaussian approximation (CG), Derivative Matching (DM) and Conditional Derivative Matching (CDM).

\subsection{Gaussian approximation}

Often, moment closure is performed by assuming a priori that the population counts have a multivariate Gaussian distribution.
Since for a Gaussian distribution all cumulants of order three and higher are equal to zero, 
moment closure function is constructed by setting the appropriate cumulant equal to zero \cite{gov06,lkk09,gou05}.  Assuming $p(t)$ and $g(t)$ are jointly Gaussian,
\begin{equation}
        \begin{aligned}
&\langle \left( g - \langle g \rangle \right) \left( p - \langle p \rangle \right)^3\rangle=\\
& \hspace{12 mm} 3 \langle \left( g - \langle g \rangle \right) \left( p - \langle p \rangle \right)\rangle   \langle \left( p - \langle p \rangle \right)^2\rangle.
        \end{aligned}
\label{eqn_fourth_second}
\end{equation}
Expanding both sides and rearranging terms, $\langle gp^3 \rangle$ can be expressed as a function of lower-order moments as follows:
\begin{equation}
        \begin{aligned}
& \langle gp^3 \rangle\approx\langle g \rangle\langle p^3 \rangle + 3 \langle gp^2 \rangle \langle p \rangle - 6 \langle g \rangle \langle p \rangle \langle p^2 \rangle \\
& \hspace{15 mm} - 6 \langle gp \rangle \langle p \rangle^2 + 6 \langle g \rangle \langle p \rangle^3 + 3 \langle gp \rangle \langle p^2 \rangle.
        \end{aligned}
\label{g3}
\end{equation}
\begin{table*}[!t]
\caption{Moment closure approximations of the fourth-order moment  $\langle gp^3 \rangle \approx \theta(\mu)$ as a function of moments up to order three.}
\label{table1}
\begin{center}
\begin{tabular}{cc}
\toprule

Moment Closure Method & Moment closure function $\theta(\mu)$, $\mu=[\langle g \rangle, \langle  p \rangle,\langle  gp \rangle, \langle p^2 \rangle, \langle gp^2 \rangle, \langle p^3 \rangle ]^T$ \\
 \midrule
Gaussian& $\langle g \rangle\langle p^3 \rangle + 3 \langle gp^2 \rangle \langle p \rangle - 6 \langle g \rangle \langle p \rangle \langle p^2 \rangle - 6 \langle gp \rangle \langle p \rangle^2  + 6 \langle g \rangle \langle p \rangle^3 + 3 \langle gp \rangle \langle p^2 \rangle$\\ \\
% \midrule
Conditional Gaussian& $ 3 \frac{\langle gp^2 \rangle\langle gp \rangle}{\langle g \rangle} - 2 \frac{\langle gp \rangle^3 }{\langle g \rangle^2 }$\\ \\
% \midrule
Derivative Matching& $\left(\frac{\langle gp^2 \rangle}{\langle gp \rangle}\right)^3 \left(\frac{\langle p \rangle}{\langle p^2 \rangle}\right)^3 \langle p^3 \rangle  \langle g \rangle$\\ \\
% \midrule
Conditional Derivative Matching& $\left(  \frac{\langle  gp^2 \rangle}{\langle gp \rangle} \right)^3 \langle g \rangle$\\
\bottomrule
\end{tabular}
\end{center}
\end{table*}

\subsection{Conditional Gaussian approximation} 

Since $g(t)$ is a Bernoulli random variable, assuming $p(t)$ and $g(t)$ to be jointly-Gaussian is quite unrealistic. Perhaps, a better approximation would be assume that  $p \mid  g=1$ (protein level conditioned
on gene being active) is Gaussian. Any higher-order moment of the form $\langle gp^j \rangle$ can be expressed in term of the conditional moment $\langle p^j \mid  g=1 \rangle$ as follows
\begin{align}
& \langle gp^j \rangle =\langle p^j \mid  g=1 \rangle \langle g \rangle, \ \ j \in \{1,2,\ldots\}\label{cond}.
\end{align}
If random variable $x$ has a Gaussian distribution, then 
\begin{align}\label{cg-zero}
&\langle \left( x - \langle x \rangle \right)^3\rangle=0 \Rightarrow \langle x^3 \rangle = 3\langle x^2 \rangle  \langle x \rangle - 2\langle x \rangle^3.
\end{align}
Assuming $p \mid  g=1$ is Gaussian, then from \eqref{cg-zero}
\begin{equation}
        \begin{aligned}
&\langle p^3 \mid  g=1 \rangle = 3 \langle p^2 \mid  g=1 \rangle \langle p \mid  g=1 \rangle - 2\langle p \mid  g=1 \rangle^3.
\end{aligned}
\label{eqn2.qo}
\end{equation}
Multiplying \eqref{eqn2.qo} with $\langle g \rangle$ and using \eqref{cond} we obtain 
\begin{equation}
\begin{aligned}
& \langle gp^3 \rangle \approx 3 \frac{\langle gp^2 \rangle\langle gp \rangle}{\langle g \rangle} - 2 \frac{\langle gp \rangle^3 }{\langle g \rangle^2}.
\end{aligned}
\label{eqn_cg3}
\end{equation}

\subsection{Derivative-matching}

At low protein copy numbers, distributions become skewed and deviate significantly from a Gaussian distribution. Not surprisingly, closure techniques based on a Gaussian distribution fail in this regime and sometimes
yield negative moments \cite{sih10
}, which are not biologically meaningful as populations cannot drop below zero. To circumvent this problem, recent work has proposed the \emph{Derivative Matching (DM)} moment closure technique, where $ \theta(\mu)$ is obtained by matching time derivatives of the exact moment equations \eqref{comp} with that of the approximate moment equations \eqref{compa}  for some initial time  \cite{sih10
}. 
Moment closure functions obtained from DM are consistent with population counts being jointly lognormal and work well in both high and low copy number regimes \cite{sih10
}. 

Theorem 1 in \cite{sih10
} provides formulas to express any given higher-order moment as a function of lower-order moments based on DM. Using this theorem, 
$\langle gp^3 \rangle$ is approximated as follows
\begin{align}\label{dm_3}
& \langle gp^3 \rangle \approx\left(\frac{\langle gp^2 \rangle}{\langle gp \rangle}\right)^3 \left(\frac{\langle p \rangle}{\langle p^2 \rangle}\right)^3 \langle p^3 \rangle  \langle g \rangle.
\end{align}

\begin{figure*}[!thpb]
      \centering
      \framebox{\includegraphics[scale=.80]{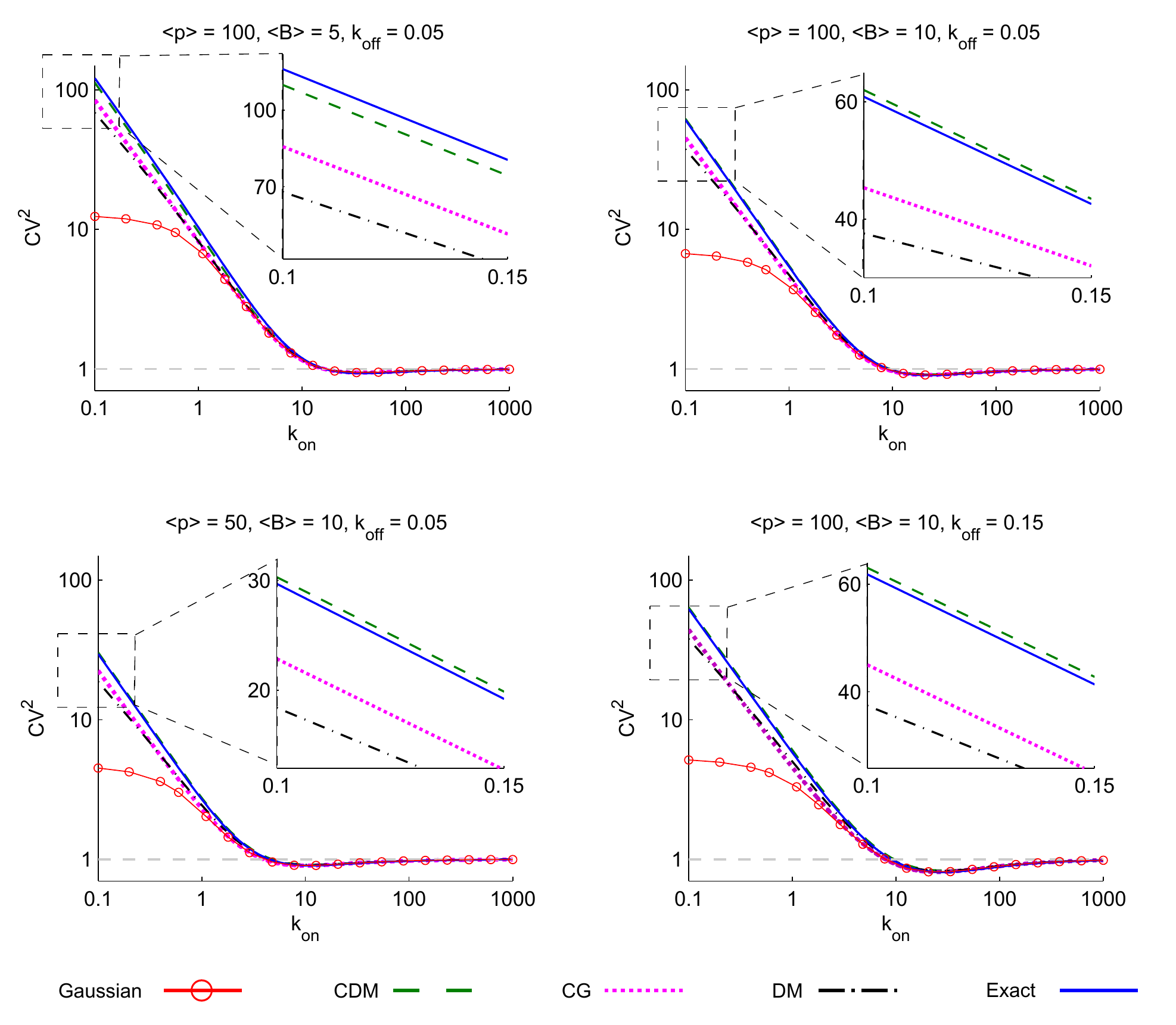}}
      \caption{{\bf Conditional Derivative Matching (CDM) moment closure technique provides the most accurate estimate of noise levels}. Protein noise levels measured by the steady-state Coefficient of Variation (CV) squared as a function of $k_{on}$ based on the exact solution, and different closure techniques (Eq. \eqref{compa} with $ \theta(\mu)$ from Table \ref{table1}) . As $k_{on}$ is decreased, $k_p$ (the protein burst arrival rate) is increased so as to keep the mean protein level $\langle p \rangle$ fixed. Subplots correspond to different values of $\langle p \rangle$, $k_{off}$ and mean burst size $\langle B \rangle$. Zoomed insets show a remarkable match between the exact noise level and those obtained  from CDM. $k_{on}$ and $k_{off}$ rates are normalized with respect to the protein decay rate $\gamma_p=1$. Noise levels are normalized their values when $k_{on}$ is large (i.e., no feedback scenario). }
\label{figresults}
   \end{figure*}
\subsection{Conditional derivative matching}
Based on DM moment closure technique, third-order moment of any random variable $x$ can be expressed as
\begin{align}\label{dm_x}
\langle x^3 \rangle =\left(\frac{\langle x^2 \rangle}{\langle x \rangle}\right)^3
\end{align}
\cite{sih10
}. Motivated by the conditional Gaussian approximation, we close the third-order conditional moment $\langle p^3 \mid  g=1\rangle$ as a function of lower-order conditional moments based on DM. From \eqref{dm_x}
 \begin{align}
\langle p^3 \mid  g=1 \rangle &=\left(  \frac{\langle p^2 \mid  g=1 \rangle}{\langle p \mid  g=1 \rangle} \right)^3 \\
  \Rightarrow \langle p^3 \mid  g=1 \rangle  \langle g \rangle & = \left(  \frac{\langle p^2 \mid  g=1 \rangle  \langle g \rangle}{\langle p \mid  g=1 \rangle   \langle g \rangle} \right)^3 \langle g \rangle,
\end{align}
which using \eqref{cond} can be written as 
\begin{equation}
        \begin{aligned}
 \langle gp^3 \rangle \approx \left(  \frac{\langle  gp^2 \rangle}{\langle gp \rangle} \right)^3 \langle g \rangle.
\end{aligned}
\label{eqn_CDM3}
\end{equation}
We refer to this approach as the Conditional Derivative Matching (CDM) moment closure technique.

Different moment closure functions derived in this section are summarized in Table \ref{table1}. Note that the conditional derivative matching technique yields the simplest form for
$\theta(\mu)$. 

\section{Comparison with exact solution}

To benchmark different moment closure methods, protein noise levels are computed from \eqref{compa} and compared to their exact values in \eqref{act}. Noise is quantified by the steady-state \emph{Coefficient of Variation (CV) squared} (variance/mean$^2$) of $p(t)$. Recall that when $k_{on}$ (gene activation rate) is large, there is no negative feedback. Decreasing $k_{on}$ implies stronger binding of protein molecules to the promoter, and is analogous to increasing the feedback strength. To understand how feedback strength affects stochasticity in protein copy numbers, we investigate steady-state protein $CV^2$ as a function of $k_{on}$. Mean protein level is made invariant of $k_{on}$ by appropriately modulating the protein burst arrival rate $k_p$. In particular, 
as we decrease $k_{on}$,  $k_p$ is simultaneously increased as per the equation
\begin{align}
k_p=\frac{\langle p \rangle\gamma_p (k_{off}\langle p \rangle + k_{on})}{k_{on} \langle B \rangle}.
\end{align}
This ensures that the steady-state protein level in the deterministic chemical rate equation model is fixed at $\langle p \rangle$.

Equation \eqref{act} yields the exact protein noise level as
\begin{align}
CV^2&=\frac{\langle p^2 \rangle-\langle p \rangle^2}{\langle p \rangle^2}\\
&=\frac{\frac{d^2G(z)}{dz^2}|_{z=1}-\frac{dG(z)}{dz}|_{z=1}-\left(\frac{dG(z)}{dz}|_{z=1}\right)^2}{\left(\frac{dG(z)}{dz}|_{z=1}\right)^2},
\end{align}
where $G(z)$ is given by \eqref{exact}. Fig. 2 plots $CV^2$ as a function $k_{on}$ for different mean protein levels, bursts sizes and $k_{off}$. Intriguingly, results reveal that for a fixed mean protein abundance, noise level is minimal at an intermediate value of $k_{on}$ (Fig. 2). Intuitively, at large values of $k_{on}$, noise level is high because of no negative feedback.  At the other extreme, a low $k_{on}$ value enhances noise due to slow switching between transcriptional states. Consequently, fluctuations in protein copy numbers are minimal at an optimal negative feedback strength. 

Protein noise levels obtained from \eqref{compa} using different functions $\theta(\mu)$ from Table \ref{table1} are also plotted in Fig. 2. Interestingly, all closure methods qualitatively capture the inverted U-shape profile. Quantitatively, moment closure based on the Gaussian approximation provides the least accurate estimate of $CV^2$. The CDM moment closure method provides a remarkably close match to the exact noise level across all parameter values, even when the gene activation rate is as low as $k_{on}=0.1$ (an order of magnitude slower activation rate compared to the protein decay rate $\gamma_p=1$).

\section{CONCLUSIONS}

Genes often employ negative feedback loops to minimize fluctuations in protein levels due to the inherent stochastic nature of gene expression. Given that enhanced stochasticity in protein levels
is associated with diseased states, developing approximate methods for studying noise buffering functions of negative feedback circuits is of considerable interest. 

Under the assumptions of short mRNA half-life and non-cooperative feedback, exact formulas for the protein statistical moments were derived. These formulas revealed an interesting fact: for fixed mean protein level, steady-state protein $CV^2$ (noise) is minimal at an optimal negative feedback strength. The exact solution for the protein moments was used to benchmark the performance of four different moment closure schemes. Our results showed large errors between the approximated and exact moments for closure based on a Gaussian approximation. Other methods (derivative matching and conditional Gaussian approximation) also showed significant errors in certain parameter regimes (Fig. 2). Our study highlights a new closure scheme, CDM, which expresses conditional higher-order moments as a function of conditional lower-order moments using the recently proposed derivative matching technique. Protein noise level obtained from CDM was an almost perfect match
with the exact solution for a wide range of parameters tested. 

Future work will investigate the performance of moment closure methods for cooperative negative feedback where $n>1$. Since the CME is analytically intractable in this case, moments obtained from running a large number of Monte Carlo simulations will be used to test the performance of closure methods. It will be interesting to see how the noise profiles in Fig. 2 change for cooperative negative feedback loops, and if there exists an optimal feedback strength where noise level is minimal.

\addtolength{\textheight}{-12cm}   % This command serves to balance the column lengths

%%%%%%%%%%%%%%%%%%%%%%%%%%%%%%%%%%%%%%%%%%%%%%%%%%%%%%%%%%%%%%%%%%%%%%%%%%%%%%%%

%%%%%%%%%%%%%%%%%%%%%%%%%%%%%%%%%%%%%%%%%%%%%%%%%%%%%%%%%%%%%%%%%%%%%%%%%%%%%%%%

%%%%%%%%%%%%%%%%%%%%%%%%%%%%%%%%%%%%%%%%%%%%%%%%%%%%%%%%%%%%%%%%%%%%%%%%%%%%%%%%
\section*{ACKNOWLEDGMENT}
\thanks{AS is supported by the National Science Foundation Grant DMS-1312926, University of Delaware Research Foundation (UDRF) and Oak Ridge Associated Universities (ORAU).}\\

\bibliography{thesis}
\bibliographystyle{IEEEtran}

%%%%%%%%%%%%%%%%%%%%%%%%%%%%%%%%%%%%%%%%%%%%%%%%%%%%%%%%%%%%%%%%%%%%%%%%%%%%%%%%

\end{document}